\documentclass[conference]{IEEEtran}
\usepackage{cite}
\usepackage[cmex10]{amsmath}
\usepackage{amssymb}
\usepackage{graphicx}
\usepackage{epstopdf}
\usepackage{graphics}
\usepackage{pdfpages}
\usepackage{float}
\usepackage{changes}
\usepackage{multirow}
\usepackage{subfigure} 
\usepackage[cmex10]{amsmath}
\usepackage{epsfig} %,subfig,algorithm,algorithmic
\usepackage{stfloats}

\begin{document}
\title{Secure Vehicular Communication: An Intelligent Physical Layer Approach}

\author{\IEEEauthorblockN{Haji M~Furqan\IEEEauthorrefmark{1}, ~Muhammad~Sohaib~J.~Solaija\IEEEauthorrefmark{1}, Jehad~M.~Hamamreh\IEEEauthorrefmark{2}~and~H{\"u}seyin~Arslan\IEEEauthorrefmark{1}\IEEEauthorrefmark{3}}

\IEEEauthorblockA{\IEEEauthorrefmark{1}Department of Electrical and Electronics Engineering, Istanbul Medipol University, Istanbul, 34810 Turkey}
\IEEEauthorblockA{\IEEEauthorrefmark{2}Department of Electrical and Electronics Engineering, Antalya Bilim University, Antalya, Turkey}

\IEEEauthorblockA{\IEEEauthorrefmark{3}Department of Electrical Engineering, University of South Florida, Tampa, FL, 33620, USA}
\\
Email:\{haji.madni, muhammad.solaija\}@std.medipol.edu.tr, jehad.hamamreh@antalya.edu.tr, huseyinarslan@medipol.edu.tr}

\maketitle
\begin{abstract}
Intelligent transportation systems (ITS) with advanced sensing and computing technologies are expected to support a whole new set of services including pedestrian and vehicular safety, internet access for vehicles, and eventually, driverless cars. Wireless communication is a major driving factor behind ITS, enabling reliable communication between vehicles, infrastructure, pedestrians, and networks, generally referred to as vehicle-to-everything (V2X) communication. However, the broadcast nature of wireless communication renders it prone to jamming, eavesdropping, and spoofing attacks which can adversely affect ITS. Keeping in view this issue, we propose the use of an intelligent physical layer security framework for V2X communication, called intelligent V2X security (IV2XS), to provide a reliable and robust solution capable of adapting to different conditions, scenarios and user requirements. Moreover, the factors affecting V2X security are identified and challenges in realizing a practical IV2XS system are discussed.

% We also identify the conditions that impact security and describe the open challenges in achieving a realistic IV2XS system.
\end{abstract}
\IEEEpeerreviewmaketitle
% \vspace{-6pt}
\section{Introduction}
The last decade has seen a significant rise in the hype regarding autonomous vehicles and intelligent transportation systems (ITS). The prime motivation behind these concepts is to handle the ever-increasing number of road accidents (World Health Organization reported 1.25 million fatalities in 2013, and millions of serious injuries \cite{WHO_traffic_deaths}), and decrease the traffic congestion, to ensure more efficient and safer mobility. Other advantages of ITS include the reduction of carbon dioxide emissions and a decrease in fuel usage which consequently helps in the conservation of non-renewable fossil fuels. The major driving technologies behind ITS include artificial intelligence (AI), sensor networks, control systems, and communication networks. The latter component arguably enjoys the utmost importance since it enables the coordination of the vehicle with all systems, either on-board, in the environment, or located at a central controlling entity.

The generic term given to vehicular communication is vehicle-to-everything (V2X), which incorporates the vehicle's communication with the network (V2N), other vehicles (V2V), infrastructure (V2I), and even pedestrians (V2P). A V2X system is expected to communicate with other vehicles and infrastructure to ensure road safety and optimize the traffic flow. %On the other hand, V2N can be used to provide internet connectivity to the users.
This necessitates a suitable communication system that takes into consideration all the possible components that need to communicate, the operating environment, possible use cases, and individual user/application requirements \cite{v2xlte}.

The traditional approach to vehicular communication uses a technology known as dedicated short-range communications (DSRC), based on IEEE 802.11p standard \cite{v2xlte}. DSRC supports communication between transmitters aboard vehicles and infrastructural road-side units (RSUs). While it works efficiently for V2V and limited V2I scenarios, DSRC lacks the scalability for large scale V2I or V2N communications due to the lack of centralized control, causing severe service degradation in congested scenarios \cite{8300313}. On the other hand, cellular-based V2X (C-V2X) boasts the ability to support larger coverage with a well-developed infrastructure but the current (till LTE-Advanced) C-V2X solutions are primarily hindered by latency. However, the fifth-generation (5G) of wireless communications promises to provide ubiquitous connectivity to all kinds of users, be it humans or machines, with stringent reliability and latency requirements. With the supported latency of 1ms for ultra-reliable low latency (uRLLC) service, this drawback of C-V2X can be addressed. This, combined with the limitations of DSRC, has led to a significant amount of research in presenting 5G-V2X as a promising candidate for future vehicular communications.

While the primary objective of C-V2X (or any other vehicular communication standard) is to provide seamless connectivity between different vehicular and infrastructural nodes, the broadcast nature of communication renders it susceptible to different attacks. Given the industrial trend in ITS regarding the increased use of AI, the need for secure communication is even more pronounced.
%While the provision of seamless connectivity is the primary objective of a communication system, the importance of its security can not be ignored, particularly in the digital age. In an AI-based use case like self-driving cars, security becomes paramount since a breach can have significant repercussions, with limited or no human supervision serving as a fail-safe backup to the machine. 
V2X systems need to ensure security against malicious attacks targeting system performance, user, and data privacy. These attacks range from stealing user information to manipulating communication in a destructive manner. User identification and authentication are particularly sensitive since they determine the legitimate access of users to data/services \cite{8300313}. %Owing to the broadcast nature of wireless communications, its security is a significant challenge and has resulted in considerable research targeted at resolving this problem \cite{5751298}.

The prevalent wireless communication security techniques can be categorized into \textit {cryptographic} and \textit{physical layer security (PLS)} methods, where the former includes key-based approaches usually applied at higher network layers but the key management has become more challenging with the heterogeneous network deployment in 5G networks. This gave rise to PLS techniques that utilize the unique properties of wireless communication, i.e., channel, interference, and noise to provide security to the users \cite{5751298}. 

The applicability of PLS techniques in the context of V2X communication has recently been discussed in \cite{elhalawany2019physical} and \cite{luo2020physical}. However, these works primarily consider eavesdropping attacks, where the illegitimate node is only interested in listening to the communication between the legitimate ones. Furthermore, integration of PLS with non-orthogonal multiple access is considered in the former work owing to the potentially large number of interaction nodes. Luo \textit{et. al.} \cite{luo2020physical} consider radio resource management methods, cooperative jamming, multi-antenna schemes, and key-based PLS methods to protect communication against eavesdropping. The authors also point out the inability of any single PLS scheme to provide appropriate security for different applications and scenarios, suggesting cooperative use of different PLS methods.

This work, driven by the above-mentioned motivation, envisions providing robust security solutions against eavesdropping, jamming, and spoofing attacks using an intelligent engine. This engine utilizes information about the radio environment and application requirements to provide the best possible PLS solution in a proactive manner, satisfying the user's security needs. The contributions of this work are listed below:

\begin{itemize}
\item  An adaptive, proactive, and intelligent security framework in V2X, referred to as intelligent V2X security (IV2XS), focusing on PLS is proposed. 

\item The factors and conditions affecting the IV2XS framework such as environment, speed, application, etc. are also elaborated.

\item An illustrative example of the concept is provided along with the challenges and open issues related to the proposed framework.
\end{itemize}

%\textcolor{green}{In this article, we first provide a quick overview of components of V2X communication and the corresponding channel characteristics in Section \ref{sec:V2X_channel}. This is followed by a brief description of security threats and possible PLS solutions in Section \ref{sec:threats}. Section \ref{sec:intelligent_security} presents the motivation and design of IV2XS and the factors affecting V2X security. The proposed IV2XS framework is then illustrated with some examples in Section \ref{sec:illustration}. Section \ref{sec:open_issues} sheds light on the open issues of IV2XS before the conclusion in Section \ref{sec:conclusion}.}
An overview of the components of V2X communication is provided in Section \ref{sec:V2X_channel}, followed by a description of security threats and possible PLS solution in Section \ref{sec:threats}. IV2XS design is discussed in Section \ref{sec:intelligent_security} while an illustrative example for the proposed framework is provided in \ref{sec:illustration}. Section \ref{sec:open_issues} sheds light on the open issues of IV2XS before the conclusion in Section \ref{sec:conclusion}.

\section{V2X Characteristics}
\label{sec:V2X_channel}
\begin{figure}[t]
        \centering
        \includegraphics[scale=0.3]{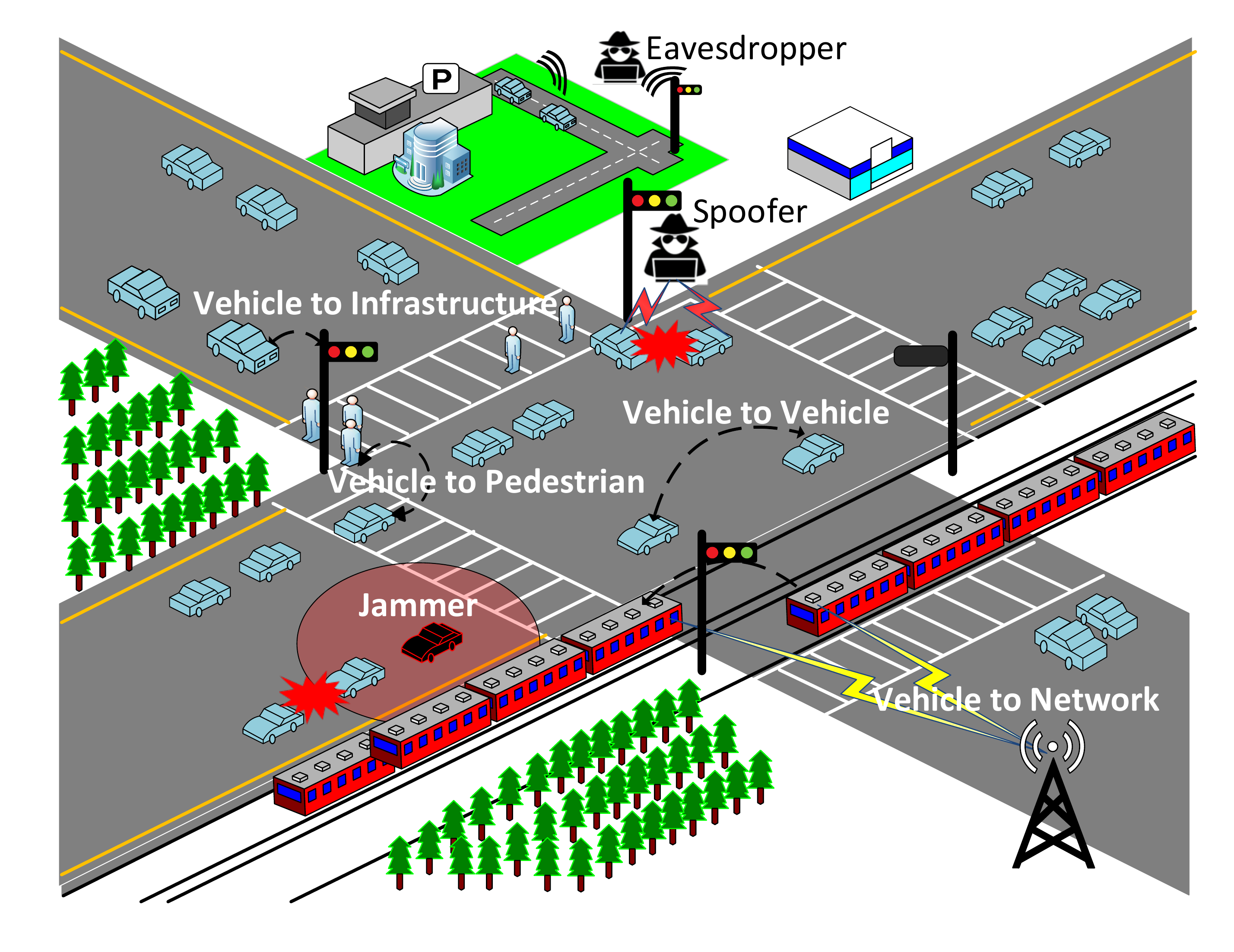}
        \caption{An overview of a V2X system and possible security threats. %V2I links are used for traffic management, V2N provides internet access to users, V2V enables collision avoidance and V2P provides safety alerts to pedestrians. Jamming is shown to disrupt V2V communication, while spoofing can manipulate the V2I interaction to cause accidents. An eavesdropper is shown to capture vehicle information at a parking.
        }
        \label{fig:V2x_overview}
    \end{figure}

V2X is an umbrella term for vehicular communications and it involves a vehicle's communication with different classes of components, all of which have their own corresponding applications. Figure \ref{fig:V2x_overview} gives an overview of a V2X system involving V2I links for traffic management, V2N for internet access to the users, V2V for collision avoidance and V2P for providing safety alerts to pedestrians and cyclists.

Multiple communication standards have been developed to ensure interoperability in information exchange between vehicles \cite{v2xlte}, however, for this work we keep ourselves focused on C-V2X because it supports both direct communication as well as communication over a cellular network. The former is carried over the PC5 interface and it includes V2V, V2I, and V2P operating in ITS bands. It is particularly suitable for latency-critical applications concerned with safety and reliability. The latter caters to V2N which uses a traditional mobile broadband licensed spectrum. V2N is concerned with latency tolerant use cases, such as telematics or infotainment.

%\subsection{V2X Channel Characteristics and 5G}
In addition to the various components, V2X also has a channel unlike any other communication application. The fundamental difference is the extremely low temporal correlation of the channel due to a rapidly changing environment, which is a consequence of the continuous vehicular mobility and the resulting high Doppler spread. In addition to this, the mobility also causes continuously changing network topology.  

% \textcolor{green}{V2X is a particularly unique use case since it does not fall under a single 5G-defined class of service. Instead, it has some aspects of all three services \cite{8031287}. For instance, the safety-related messages of V2X have strict latency and reliability constraints, which corresponds to the uRLLC service. Infotainment and other rich data sharing applications have an element of eMBB eer number of vehicles on road relates to the mMTC service of 5G.}
%whıch three 
% The technologies touted to enable 5G-V2X communication include network slicing, supported by network function virtualization (NFV) and software-defined networking (SDN) to enable the provision of various services, in addition to edge computing enhancements which are targeted at reducing the latency of the system and multipoint cooperation which enhances the system reliability \cite{5G_CAR}. 

% \vspace{-6pt}
\section{Security Threats in V2X and Solutions}
\label{sec:threats}
\begin{figure*}[t]
        \centering
        \includegraphics[scale=0.62]{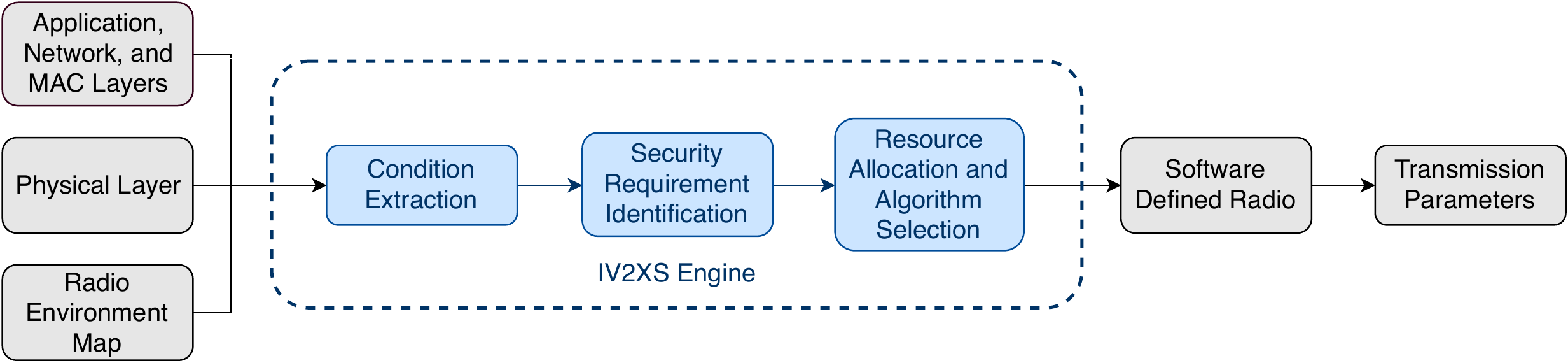}    
        \caption{Conceptual system model for the IV2XS approach. Information from REM and different network layers is exploited by the AI-based engine to allocate proper security resources and algorithms using the SDR platform.}
        \label{fig:system_model}
    \end{figure*}
\subsection{Security Threats in V2X}
The broadcast nature of wireless V2X communication makes it vulnerable to \textit{eavesdropping}, \textit{spoofing}, and \textit{jamming} attacks. These attacks can cause serious problems for V2X communication, especially for autonomous or remote driving and other critical cases in terms of safety, privacy, and efficiency. In \textit{eavesdropping}, an illegitimate receiver tries to intercept the communication between legitimate parties, thus violating confidentiality and privacy. Figure \ref{fig:V2x_overview} shows the example of an eavesdropper trying to access parking-related information of a user. In the case of \textit{jamming}, the illegitimate node generates intentional interference to disrupt the communication between the legitimate nodes. As shown in Fig. \ref{fig:V2x_overview}, a jammer might try to interrupt the communication between vehicles, forcing them to collide. Finally, in the \textit{spoofing} attack, the control of the communication channel between the legitimate parties is taken by spoofer. The spoofer can replace, modify, and intercept the messages that are being transmitted between two legitimate parties \cite{oursur1}. A spoofing attack on V2I communication might result in vehicles moving in direct collision paths of each other, as shown in Fig. \ref{fig:V2x_overview}.

There are two popular security approaches to tackle these attacks in V2X communication: cryptography-based solutions and PLS-based solutions. 
The former, based on key sharing by a trustable third party, are effective in providing secure data communication in current 5G V2X and other wireless systems \cite{hamida2015security}. However, cryptography-based solutions may not be suitable for future V2X wireless communication because the management and maintenance of keys are very challenging tasks in a decentralized and heterogeneous environment such as V2X communication. This is further compounded by intermittent connectivity and the varying speed of the V2X entities. Moreover, the security of the key-sharing process is critical, i.e., if the key is intercepted during the said process, all subsequent transmissions are liable to illegitimate access. Additionally, the sensors, actuators and transceivers utilized for instant control in autonomous or remote-controlled driving are processing restricted, power-limited, and delay-sensitive. These limitations render them incapable of supporting sophisticated encryption/decryption techniques needed for cryptographic security solutions.

In order to handle these issues, PLS techniques have emerged as an effective security solution for future V2X communication that can complement and even replace the cryptography-based approaches \cite{oursur1}. PLS exploits the dynamic features of wireless communication to secure the link between legitimate nodes. It has the following potentials as a solution for future communication security. Firstly, these approaches can extract keys from the time-varying wireless channel, avoiding key management and maintenance issues in decentralized V2X wireless networks. Secondly, they are also suitable for power-restricted and delay-sensitive applications since many of them can be implemented by relatively simple signal processing algorithms \cite{5751298}.
\subsection{PLS Solutions in V2X}
\subsubsection{Anti-eavesdropping solutions for V2X}
There are several PLS techniques in the literature against eavesdropping applicable for V2X, such as channel-based adaptation and key extraction, channel-coding design-based methods, and injection of artificial noise. The basic idea in channel-based adaptation is to modify the transmission parameters based on the requirements, location, and wireless fading channel conditions of the legitimate receiver \cite{oursur1}. Examples of this approach include beamforming, adaptive modulation and coding, directional modulation, and adaptive power allocation \cite{oursur1}. Channel-based key extraction techniques generate high-rate keys from the dynamic vehicular wireless channel \cite{7876781}. The methods based on channel coding design use special channel codes to ensure secure communication. In the case of artificial noise-based approaches, an interfering signal (noise/jamming) is added by exploiting the null space of the legitimate V2X node’s channel to degrade the performance of eavesdroppers. %An interesting example of this is the exploitation of road geometry and natural restrictions on the potential eavesdroppers' locations to provide security through injecting artificial/jamming noise in a controlled manner along the direction of travel\cite{7876781}.

    \begin{table*}[t]\centering\renewcommand{\arraystretch}{1.35}
\caption{Summary of security threats, risks and IV2XS framework examples.}
\label{table:PLS_Solutions}
\centering\resizebox{1.97\columnwidth}{!}{
\begin{tabular}{|c|l|l|}
\hline
\textbf{Threats}       & \multicolumn{1}{c|}{\textbf{Risks}}                                                                   & \multicolumn{1}{c|}{\textbf{IV2XS Framework Examples}}                                                                                                                                            \\ \hline
\textbf{Eavesdropping} & \begin{tabular}[c]{@{}l@{}}Stealing of personal,  financial,\\  and location information.\end{tabular}                         & \begin{tabular}[c]{@{}l@{}}\textbf{High security}: Adaptive artificial noise based,  interference-based, pre-coding, and hybrid techniques.\\   \textbf{Medium security}: Beamforming, directional modulation and pre-coding techniques.\\   \textbf{Low security}: Interleaving, adaptive modulation and coding.\\\end{tabular}                                                                                   \\ \hline
\textbf{Spoofing}      & \begin{tabular}[c]{@{}l@{}}Car hijacking and stealing, \\  traffic  disturbance, accidents,\\  fake messages.\end{tabular}    & \begin{tabular}[c]{@{}l@{}}\textbf{High security}: Joint features extraction from CSI,  RSS and AFE imperfections for authentication.\\   \textbf{Medium security}: More than one feature extraction from CSI/RSS/AFE imperfections.\\  \textbf{Low security}:  Single feature extraction from CSI/RSS/AFE imperfections.\end{tabular} \\ \hline
\textbf{Jamming}       & \begin{tabular}[c]{@{}l@{}}Traffic  disturbance, accidents, \\  communication disturbance,  \\  loss  of control.\end{tabular} & \begin{tabular}[c]{@{}l@{}}\textbf{High security}: Multi-antenna approaches,  SS with more processing gain.\\   \textbf{Medium security}: SS with medium processing gain, relays.\\   \textbf{Low security}: Relays, SS with less processing gain.\end{tabular}                                                                                                            \\ \hline
\end{tabular}}
\end{table*}
\subsubsection{Anti-spoofing solutions for V2X}
The authentication based on conventional cryptography is a significant challenge for V2X communication networks. This is due to the unwanted latency caused by complex backhaul processing and multiple handshakes between users, base stations (BSs), and roadside units (RSUs) for pairwise key or information exchange. However, message or entity authentication in a highly dynamic vehicular communication system can be done in a faster and more robust way using PLS approaches \cite{7498103}. For example, reciprocal channel properties such as channel state information (CSI) and received signal strength (RSS) between two communicating entities of V2X system, or the analog front-end (AFE) imperfections of wireless V2X transceivers such as carrier frequency offset (CFO) and in-phase/quadrature imbalance (IQI) can be exploited to authenticate the communicating nodes. %Moreover, the speed, location, and direction of vehicles can also be used for this purpose.

\subsubsection{Anti-jamming solutions for V2X}
Jamming attacks can cause serious problems for V2X scenarios by interrupting legitimate communication between different nodes, leading to traffic disruption or accidents. There are three broad categories of PLS-based solutions against jamming attacks: multi-antenna based approach, which is an effective solution because of its ability to avoid interference from unwanted sources \cite{kosmanos2016mimo}; cooperative relaying schemes, where V2X entities can act as relays, have the ability to re-route the traffic \cite{8336901}; and spread spectrum (SS) techniques, (e.g. direct sequence spread spectrum (DSSS) and frequency-hopping spread spectrum (FHSS)), which can also be used by V2X entities against jamming attacks by spreading the signal or by rapid frequency switching \cite{5751298}.

% \vspace{-7.5pt}
\section{Intelligent Security Design for V2X Communication}
\label{sec:intelligent_security}
The security needs vary dramatically depending on the environment/medium, scenario/use-case, and application/service associated with each legitimate transmission. This necessitates the use of intelligent security design.
\subsection{Intelligent Security Design}
%5G V2X communication is not encompassed by any single service of 5G, rather it contains the flavors of each of eMBB (infotainment), uRLLC (collision avoidance) and mMTC (large number of communicating nodes on the road). In other words, it means that throughput, reliability, latency, and coverage demands are not consistent for different aspects of V2X communications. Additionally, the 
The different components of V2X may have different uses, ranging from collision avoidance (V2V) to monetary transactions such as toll payment (V2I) to onboard entertainment (V2N). It seems intuitive that the criticality of security is in descending order in the above-mentioned tasks. This is just a simple example illustrating the varying requirements of a few different use cases concerning various V2X components. If the entire V2X communication was limited to this (i.e., different components having their own level of security requirements), the provision of adaptive security related to particular components would have been relatively easy. This, unfortunately, is not the case. The required security levels depend not only on the components that are communicating but also on the particular application, location of the user, utility, environment, etc. To cater to this, V2X communication security needs an intelligent framework.
% width=0.85\columnwidth
\begin{figure*}[t]
    \centering
    \subfigure[]{
        \includegraphics[scale = 0.27]{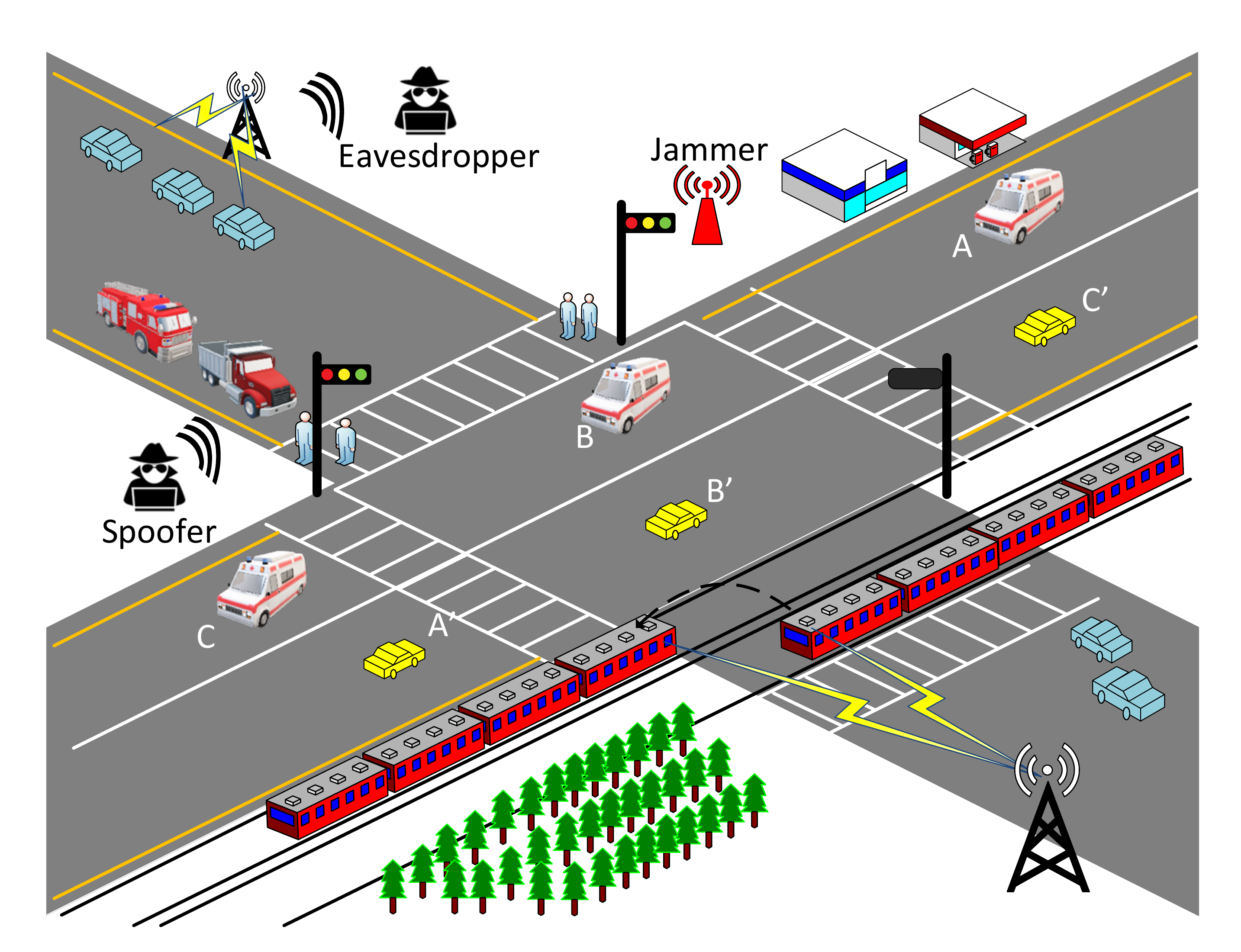} \label{fig:conditions}
    }    
    \subfigure[]{    
        \includegraphics[scale = 0.48]{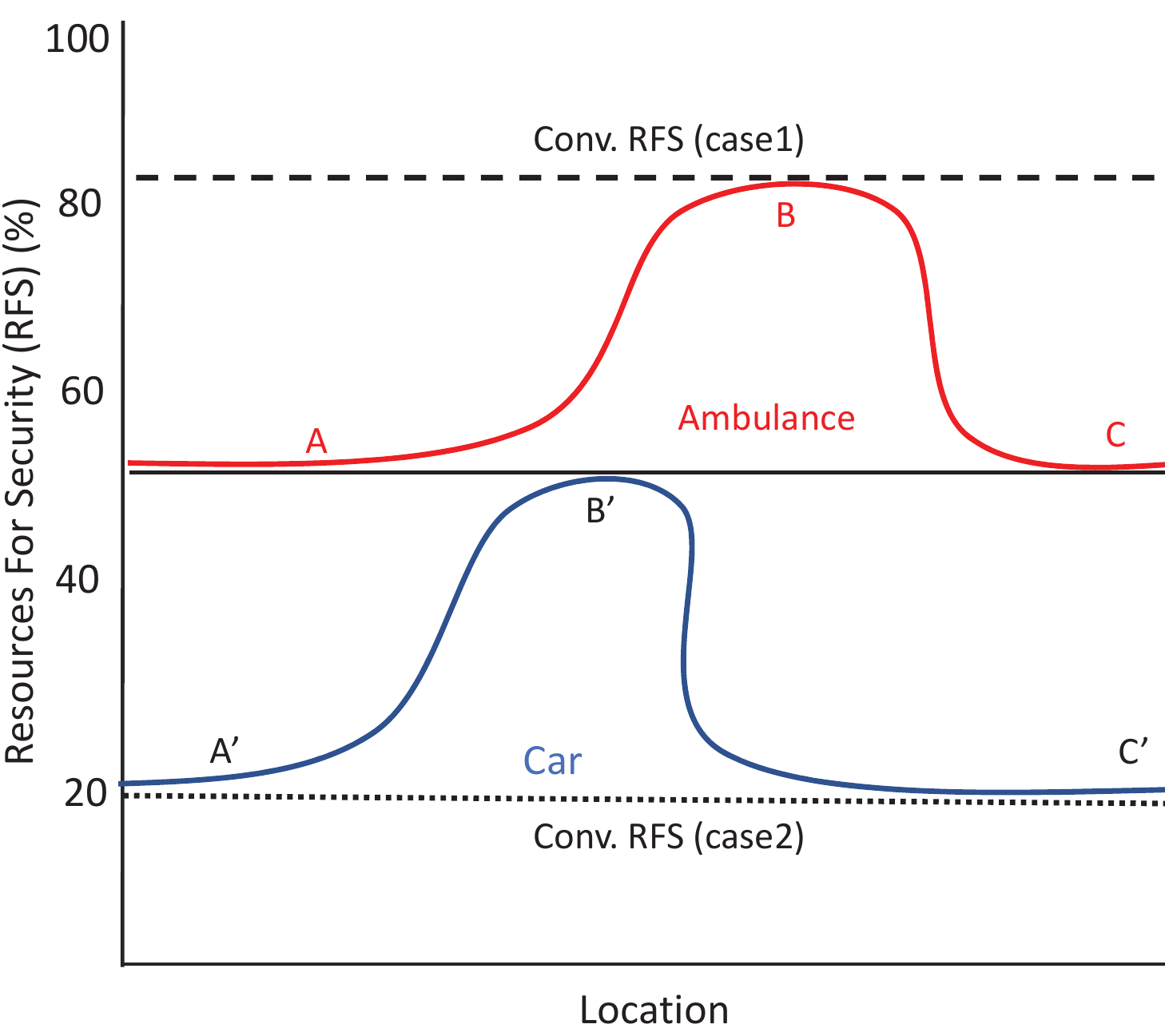} \label{fig:examples}
    }
    \caption{(a) An illustrative example regarding different conditions and security requirements, the vehicles being considered are the ambulance and the yellow car (b) Security resource allocation corresponding to the scenario and vehicle locations (A-B-C for an ambulance, A'-B'-C' for the yellow car) in (a).}
    \label{fig:illus2}
\end{figure*}

IV2XS is driven by the idea of providing proactive, adaptive, and efficient security to V2X entities. Figure \ref{fig:system_model} illustrates the conceptual IV2XS framework. It is powered by an AI-engine with input from radio environment map (REM), physical and higher network layers. REM is a cognitive enabler that provides information such as user distribution, traffic levels, and power maps, allowing smart network planning. The application layer provides information regarding the application requirements while the physical layer provides instantaneous CSI that can be utilized for different PLS schemes. The AI-powered engine makes use of the provided information to extract conditions such as location, utility, application, environment, situation, and vehicle specifications. Different conditions necessitate varying security requirements (this is elaborated in section \ref{subsec:conditions}). Depending on these requirements, appropriate resources and methods from a pre-defined set will be allocated and then provided using a software-defined radio for ensuring secure communication.

Table \ref{table:PLS_Solutions} presents the different types of potential attacks on V2X communication and their possible solutions by IV2X engine based on the security requirements. The requirement of security can be variable based on the criticality of the conditions. For the purpose of this paper and the illustrations within, we have considered three security levels, low, medium, and high. Here, low may correspond to illegitimate access to user information such as the entertainment content being accessed while high-security level might refer to the security of an emergency vehicle at a busy junction.

\subsection{Conditions Affecting IV2XS Security Framework}
\label{subsec:conditions}
\subsubsection{Location}
Security threats and V2X entity locations have a high degree of correlation. In essence, the location where a security breach can affect more users requires stronger protection. Consider the case where two V2X entities are following each other on an otherwise deserted road, they only need to know about the other vehicle's speed and distance to keep a safe cushion to avoid any collision. Now take the case of two vehicles at an intersection, there is an increased number of factors that affect the decision making process for vehicles about when to move and in which direction, resulting in higher security requirements. In addition to monitoring each other, these vehicles also need to take into account the traffic movement in other directions, which is dictated by the traffic signals. A spoofing attack, in this case, can wreak havoc by making vehicles from different directions move at the same time.

To cater to such situations, IV2X can be used to raise the level of security once it detects the location to be an intersection, ensuring nothing untoward happens. Figure \ref{fig:illus2} shows the variation in security requirements as a function of the location. The positions of two vehicles, ambulance, and the yellow car are labeled as points A-B-C and A'-B'-C', respectively. Here B and B' represent the vehicles in the middle of the intersection, where the security requirement as well as allocated resources for it increase, as shown in Fig. \ref{fig:examples}, to account for the higher risk. Security is also critical in the case of mountains or bridges or generally any location that can act as a bottleneck for traffic. On the other hand, the required security level inside gated communities or university campuses is relatively low.

% \vspace{-2pt}
\subsubsection{Utility}
Another factor that determines the necessary level of security is the type of V2X entity and its usage. A vehicle may be private (belonging to individual or companies), public (public transportation, municipality vehicles), or belong to emergency services like police, ambulance, or fire brigade. It stands to reason that in the case of an emergency the latter category of V2X entities should be given priority on the road and in the communication. 
If we consider a successful breach, it is logical that a breach targeting an emergency vehicle can cause more damage than that of an individual's vehicle, which means these vehicles need higher security. This is also illustrated in Fig. \ref{fig:examples} where the security level of the ambulance and the corresponding resources allocated for it are higher than a regular car (shown in yellow in Fig. \ref{fig:conditions}). %Similarly, in the case of platooning, i.e., when a group of vehicles coordinate with each other and move closely at high speed, with a lead vehicle that decides the speed and direction of movement, security is very critical. In such a case, the security level of the leader of the platoon is different as compared to other members. Therefore, IV2XS needs to take into consideration the purpose of the vehicle before assigning a security level to it.
\begin{figure}[t]
    \centering
    \subfigure[]{
        \includegraphics[width=0.83\columnwidth]{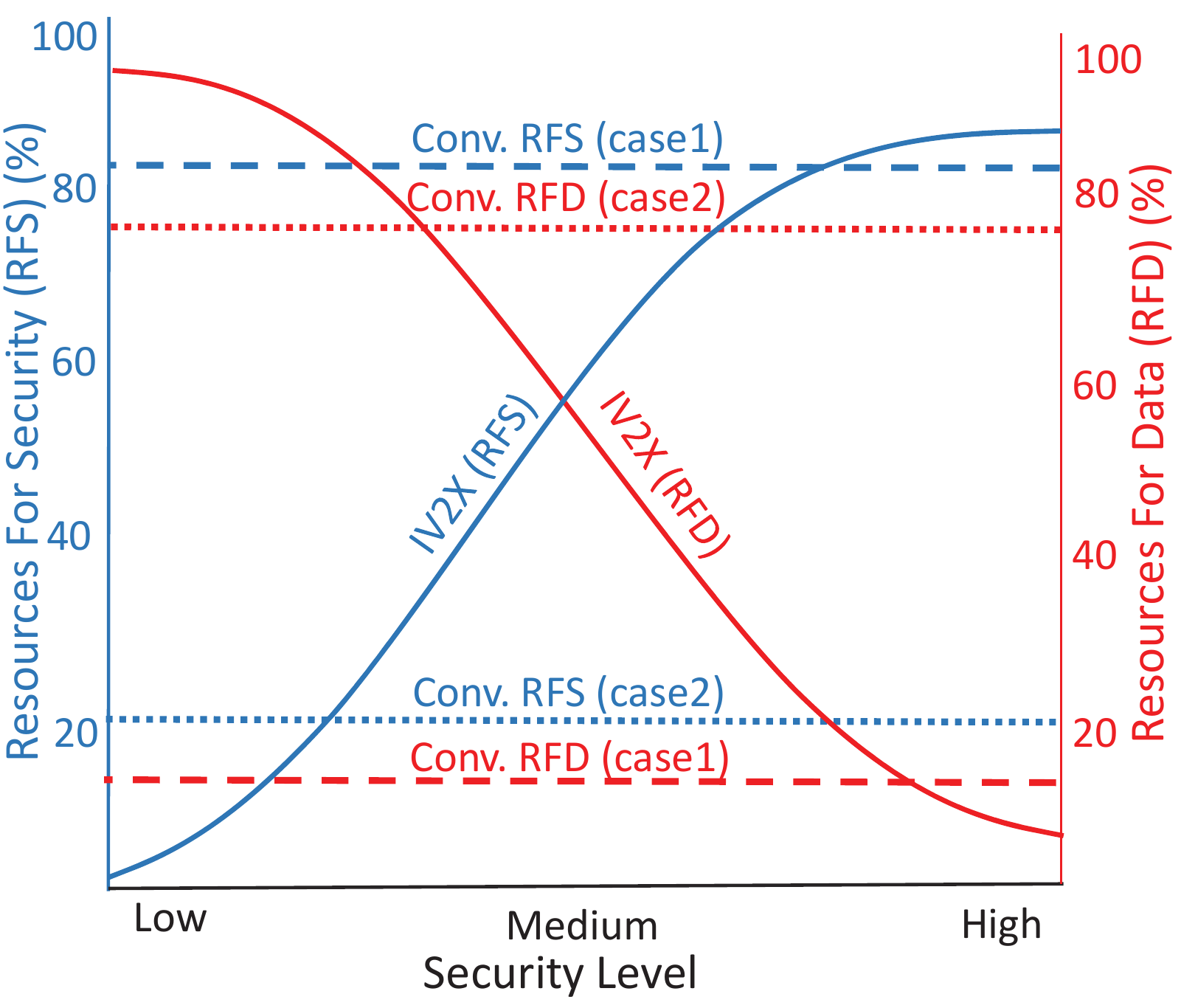} \label{fig:illus1}
    }    
    \subfigure[]{    
        \includegraphics[width=0.83\columnwidth]{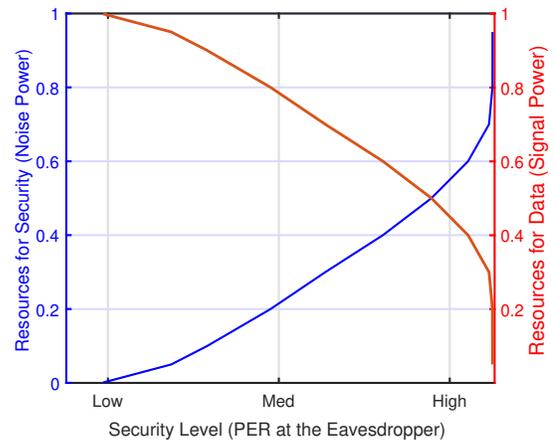} \label{fig:illus1sim}
    }
    \caption{(a) Illustration for comparison of the proposed framework versus conventional approaches, (b) Simulation results in terms of eavesdropper packet error rate (PER) vs noise power for an adaptive artificial noise based solution.}
    \label{fig:illus22}
\end{figure}

% \vspace{-2pt}
\subsubsection{Application}
From the security perspective, the application is an important parameter for the IV2XS concept. There are different types of applications in V2X communication that may be safety-related (e.g., collision avoidance and cooperative driving, queue warning), traffic-related (e.g., optimizing the traffic flow), infotainment (e.g., internet access and video streaming), payment applications (e.g., toll collection), location-based application (e.g., finding the closest fuel station). The above-mentioned applications require different levels of communication security.
Safety-related applications would require more resources for security while infotainment applications employ more resources to achieve high data rates. 

% \vspace{-2pt}
\subsubsection{Environment}
The social environment is primarily categorized into rural, suburban, and urban. These environments differ from each other on the basis of population, V2X entities density, and available infrastructure. The environment is a significant factor in deciding the security levels of V2X entities. 
The traffic density is higher in urban areas, and it is plausible to assume these areas would also be attacked more often which consequently dictates the level of security needed. The illustration in Fig.  \ref{fig:illus2} shows an urban environment with relatively high security requirements, potentially using a significant amount of the available resources for ensuring the privacy of the communication. 

\subsubsection{Situation/Time}
Existing security threats would differ based on time and the situation in which communication is taking place. The density of vehicles is different at different times. For example, vehicle density is high during the day, particularly during office hours, and generally low at night. In general, the probability of threats increases with the increase in density. The IV2XS engine will adjust the security resources accordingly. Additionally, weather conditions would also affect V2X communication requirements. For instance, the required latency of V2V messages would be different for a normal road as compared to that for a road that is slippery due to snow, which would affect the security requirements.
\subsubsection{Vehicle Specifications}
Vehicle specifications such as size, engine capacity, power, and mileage also need to be considered for security requirements. A larger vehicle can cause more harm to the vehicles or infrastructure around it and needs to have higher security than smaller ones. This is another parameter that is reflected in Fig. \ref{fig:illus2}, where the larger size (and speed) of the ambulance also contributes to higher security requirements.  
Similarly, the degree of autonomy of the V2X entity has to be taken into account as well, it follows basic intuition that a fully autonomous vehicle is more dependent on V2X communications, and therefore can be affected more severely by a breach in security.

% \vspace{-6pt}
\section{IV2XS: An Illustrative Example}
\label{sec:illustration}
Given that this article proposes a generic framework for intelligent security provision for V2X communication, an implementation or simulation covering the whole scope of the framework is relatively tedious and impractical. 

Therefore, we have taken the liberty of explaining the concept with, first, a generic comparison between conventional and proposed approaches for security provision in terms of resource allocation for data and security. Second, we consider a specific artificial noise-based adaptive technique against eavesdropping and look at the resource distribution for different levels of security.

Fig. \ref{fig:illus1} shows how the proposed approach would stack up against conventional methods in terms of resource allocation for security. In a conventional approach, it is possible to tune the network parameters either to the worst-case scenario (see the dashed lines) or an average/general case (see the dotted lines). In the former case, resources are unnecessarily allocated for security since the worst-case scenario rarely occurs, leading to scarcity of resources for the data. In the latter case, the assigned security resources are too little to thwart any strong attacks, leaving the communication vulnerable to being listened, interrupted, or manipulated. The solid lines represent the IV2XS resource division. Since it is able to adapt to the usage scenario and attack types, it provides more efficient resource utilization by only sacrificing the system capacity when absolutely needed. 

To illustrate the concept further, we give the example of an adaptive technique capable of providing security against eavesdropping for different application requirements by adjusting the power level of the artificial noise \cite{hamamreh2018joint}. Figure \ref{fig:illus1sim} shows the security level in terms of packet error rate (PER) at the illegitimate eavesdropper as the noise power is changed. The horizontal axis shows different security levels, where low security refers to 10$\%$ PER, medium refers to 50$\%$ PER and high refers to 90$\%$ PER at the eavesdropper. In this particular simulation setup, the total power (signal + noise) is kept constant, and the normalized power levels for both are shown in the figure on the vertical axes, where the blue and red lines (and axes) represent the noise and signal power levels, respectively. An alternate approach is to add noise without reducing the signal power, this would improve the legitimate user's performance as compared to the previous case at the cost of increased total transmit power. It should be kept in mind that the increased security requirement means the availability of fewer resources for data transmission, leading to degraded throughput even for the legitimate users. This necessitates the need for adaptability of the security approaches, the lack of which would adversely affect the network performance, in terms of either capacity or security. 
Here it is important to note that the proposed framework is NOT limited to a single adaptive technique, such as the one described above. Rather, it is capable of switching between different algorithms, potentially even using multiple of them simultaneously to avoid eavesdropping, jamming, and/or spoofing attacks.

% \vspace{-3pt}
\section{IV2X Challenges and Potential Solutions}\label{sec:open_issues}
% \vspace{-2pt}

%PLS techniques exploit the properties of wireless communication to provide security. One of these properties is the channel itself and acquiring its information 
\subsection{Security for High Mobility}
%\textcolor{green}{PLS techniques provide security by exploiting the properties of wireless communication, the channel being the most important one.} 
Acquiring CSI is central to several security algorithms under the PLS umbrella. However, owing to the high mobility in the V2X scenario and consequent channel variation, estimating and tracking the channel becomes very challenging. One approach to ease this burden is to exploit the channel sparsity, which means there are fewer parameters to estimate \cite{7131541}. Alternatively, PLS techniques that require partial or no CSI can be utilized, such as, directional modulation, interference-based algorithm \cite{oursur1}, etc. Yet another alternative is to use the time-invariant characteristics of the communication to provide security. For instance, radio frequency (RF) fingerprinting is an approach that leverages the uniqueness of AFE imperfections of the devices for their authentication, thwarting spoofing attacks in the process \cite{wang2016wireless}.

% \vspace{-7.5pt}
\subsection{Condition Detection}
\vspace{-3pt}
The detection of the condition information at the IV2XS engine is the first step for the IV2XS security approach. As explained earlier, in this work we are considering six conditions, i.e., location, utility, application, time/situation, environment, and vehicle specifications. The information from REM and different layers of the communication system is leveraged to identify the condition. In addition to the user and traffic densities, REM can provide information about the expected RSS levels, which can be used to localize the vehicles. Furthermore, application-specific requirements are provided to the IV2XS engine from the corresponding network layer. %Here it is important to note that the information provided by the different input elements of the IV2XS engine is not static, i.e., the conditions not only need to detected but also tracked for each user over time. 
% \vspace{-7.5pt}
\vspace{-3pt}
\subsection{Security Level Identification}
\vspace{-3pt}
Once a user's condition is determined, the next step is to define the corresponding level of security threat. One possible approach is to define three security levels: low, medium, and high, where low refers to data/information theft, medium level corresponds to possible damage to property and high-security level refers to life-threatening situations. We need to reiterate that this is not a thorough categorization of the security levels, and depending on the available resources, the security levels might be changed. While having more security levels would ensure more efficient resource utilization, it would come at the cost of increased complexity. Balancing this trade-off, or proposing an improved approach is an open research area.

% \vspace{-7.5pt}
\vspace{-3pt}
\subsection{Security Mechanism and Resource Allocation}
\vspace{-3pt}
After identifying the condition and its corresponding security level, the next step is to select suitable algorithms and allocate appropriate resources to achieve that goal. Continuing with respect to the above-mentioned example of Fig. \ref{fig:illus2}, the car at the intersection with high-security risk level, where any attack can cause a major problem, will be allocated more resources and stronger algorithms by the IV2XS engine. More specifically, the IV2XS engine can select artificial noise injection with multi-antenna based approaches at the transmitter to protect information from eavesdropping while simultaneously using interference alignment techniques at the receiver to combat jamming \cite{oursur1}. The design of artificial noise can be a function of the security threat level. On the other hand, for the low-security threat level, simple security techniques such as adaptive resource allocation based algorithms can be selected proactively by the IV2XS engine. It is also possible to use multiple security algorithms in conjunction with each other when the application requires a higher level of security. In this case, it is advisable to use an AI-based approach \cite{AI} to decide upon the most suitable algorithms and their respective resources.

% \vspace{-7.5pt}
\vspace{-3pt}
\subsection{Challenges Related to PLS}
\vspace{-3pt}
In general, PLS techniques are sensitive to channel reciprocity and estimation mismatch errors. These errors should be considered while designing future PLS techniques, and novel and effective channel estimation algorithms need to be proposed. Similarly, more efficient designs for noise-based techniques like the one described above are needed.

%Artificial noise-based PLS techniques are effective solutions for security. However, they sacrifice power resources and might cause an increase in the peak to average power ratio of the system. A redesign of noise-based techniques is needed to provide security while enhancing the performance of the system.

%Generally, in PLS techniques, transmission parameters of the physical layer are optimized according to legitimate users’ channel characteristics with the sole purpose of providing secure communication. However, joint design of security, reliability, throughput, and delay should be studied for future 5G V2X PLS techniques, considering the trade-offs among them. The concept of cross-layer security can be utilized for such joint optimization. {Cross MAC-PHY}, {NET-PHY}, and {APP-PHY} approaches use features belonging to the MAC (channel accessing, multiplexing, etc.), network (relaying, routing, etc.), and application layers (application requirements), respectively in concurrence with physical layer parameters to provide efficient QoS based security solution. Finally, hybrid techniques integrating signal security approaches with data security algorithms can provide an effective solution for V2X-based systems.

\section{Conclusion}
\label{sec:conclusion}
Security is pivotal in wireless communication, particularly in a use case like V2X that depends heavily on seamless and reliable connectivity between its different components. In this paper, we have introduced the use of an intelligent proactive framework for PLS that detects the condition of a user, considers the channel and upper-layer information before making a decision about the best-suited security level and allocating resources for ensuring secure communication accordingly. Here it is important to reiterate that the focus of this work is to provide an initial framework for IV2XS and highlight the factors affecting it, rather than focusing on specific PLS techniques. We believe the proposed intelligent framework would prove to be a stepping stone towards secure V2X communications.

% % \section{Comments}
%  \textcolor{blue}{- Contribution/motivation\\
%  - technical aspect of contribution\\
%  - related literature (PHY intelligence), PHY layer intelligence for vehicular communication\\
%  - real-time performance evaluation\\
%  - specific issue needs to be highlighted in abstract\\
%  - security problems of V2X are NOT comprehensively covered\\
%  - performance comparison of different schemes using real-time data/scenarios\\
%  - Other articles addressing same problem???}
%  - incorporating different beamforming techniques
% (i.e., analog, digital, and hybrid),

\end{document}